\documentclass[sigconf]{acmart}

\setcopyright{none}
\acmISBN{}
\acmDOI{}
\acmYear{}
\usepackage[utf8]{inputenc}
\usepackage{hyperref}
\hypersetup{
    colorlinks=true,
    linkcolor=blue,
    filecolor=magenta,      
    urlcolor=cyan,
    pdftitle={Overleaf Example},
    pdfpagemode=FullScreen,
    }
\usepackage{caption}
\usepackage{subcaption}

\title{Studying the anonymity trilemma with a discrete-event mix network simulator}
\author{Ania M. Piotrowska}
\affiliation{%
  \institution{Nym Technologies}
  \streetaddress{}
  \city{}
  \state{}
  \country{}
  \postcode{}
}

\begin{document}
\begin{abstract}\label{sbtract}

In this work, we present a discrete event mix network simulator, which allows analysing how anonymity, latency, and bandwidth overhead are affected by various scenarios of deployment and design choices. These design choices include network topology, mixing technique, volume of traffic, latency requirements, packet size or use of cover traffic. To the best of our knowledge, this is the first such simulator as work on it began in 2017 to analyze the Loopix mix network, and the code of our simulator is available under an open-source license. 
To demonstrate the capabilities of our simulator, we perform an empirical analysis of the impact of core design choices on anonymity, scalability and latency in Elixxir, HOPR and Nym, currently deployed mix network infrastructures that make a variety of different choices in their design. 

\end{abstract}

\settopmatter{printacmref=false}
\maketitle

\section{Introduction}\label{sec:introduction}

Network traffic is a source of privacy-sensitive information,
both in terms of the content and metadata.
Encryption is only the first step in the efforts to achieve privacy in the online world, since it can only protect the confidentiality of our communication
, but does not hide the unique characteristics associated with the communication, the so-called \emph{metadata}.
Exposing the metadata not only undermines the properties provided by the use of encryption~\cite{DBLP:journals/ccr/CoullD14, DBLP:conf/ndss/BahramaliHSGT20, DBLP:conf/wisa/ParkK15, DBLP:conf/uss/SongWT01, DBLP:conf/sp/DyerCRS12, DBLP:conf/www/EnglehardtREZMN15, DBLP:conf/sp/ChenWWZ10, DBLP:conf/ndss/SibyJDVT20}, but also can be exploited to uniquely identify users, enable pervasive tracking,
and reveal private information about personal or business activities~\cite{cadwalladr2018revealed, tracking2, tracking3, matz2017psychological, DBLP:journals/pnas/MayerMM16, DBLP:conf/percom/GreschbachKB12, LandauDiffie, chinasurveillance}.


Anonymous communication networks hide the network metadata and the relationship
between communicating parties on the Internet.
Since 2004, Tor has become the most popular anonymous communication tool, attracting at least 2 million users daily. However, as research has shown~\cite{DBLP:conf/wpes/PanchenkoNZE11, DBLP:conf/ndss/PanchenkoLPEZHW16, DBLP:conf/wpes/PanchenkoMHLWE17, DBLP:conf/diau/SyversonTRL00, DBLP:conf/sp/OverlierS06, DBLP:conf/sp/MurdochD05, DBLP:conf/ccs/Murdoch06} Tor offers limited security guarantees against traffic analysis techniques.
For that reason, over the past years, there has been an increased interests in \emph{mix network} technologies, with companies like Elixxir, HOPR or Nym collectively attracting millions in investment~\cite{nym_invest} with widely contrasting designs.

While great strides have been made, a key obstacle to the real world deployment of mix networks is that it is difficult to tune the trade-offs between privacy and network performance in a given design.
Therefore, we present a discrete event mix network simulator, which allows us to empirically evaluate the privacy and performance properties of various mix network designs via simulation.
In particular, our simulator allows analysing how anonymity, latency, and bandwidth overhead are affected by various scenarios of deployment of such networks and design choices.
We next use the presented simulator to investigate how anonymity and end to end latency are affected by user loads, network topology, and mixing strategies on the example of Elixxir~\cite{elixxir}, HOPR~\cite{hopr} and Nym~\cite{nym}. To the best of our knowledge, this is a first such empirical analysis of those networking designs.


\section{Background}\label{sec:background}

The concept of an anonymous communication network was pioneered by Chaum~\cite{DBLP:journals/cacm/Chaum81} to mitigate the threats of progressive communications surveillance. Their main security goal is to shield the correspondence between senders and recipients by hiding the distinctive characteristics of the network traffic, the so called \emph{metadata}.
The principal concept of the Chaum’s anonymous communication network is the decentralized \emph{mix network} in which the network traffic is relayed via a set of independently operated cryptographic relays called \emph{mix nodes}.
In contrast to onion routing~\cite{DBLP:journals/cacm/GoldschlagRS99}, however, mixnets route each individual packet via independent routes. By routing the traffic through a sequence of  independent nodes, mix networks protect users against disclosing their identity to end receivers.

In order to ensure anonymous communication, the traffic is layer encrypted using public-key cryptography.
The encapsulation of the packets into a special cryptographic packet format provides the \emph{bitwise unlinkability} between the bit patterns of the encoded messages entering and decoded ones leaving the mix node. This ensures that any third-party observer is not able to trace the packets based on their binary representation. Moreover, malicious nodes which route the packets cannot eavesdrop on user traffic, as none of the mix nodes knows the whole connection, they only know the previous and next hop, therefore only the first mix knows who is the sender, and the last mix knows who is the recipient.

In addition to transforming the routed packets cryptographically, the nodes perform the reordering (\emph{mixing}) operation so that the incoming and outgoing packets cannot be linked based on timing. Mixnets thus provide strong metadata protection against an adversary that can observe the entire network and perform traffic analysis attacks.

Originally, Chaum proposed to relay all packets through a fixed set of mix nodes, called a \emph{cascade}. The mixes used the \emph{batch-and-reorder} mixing technique,  in which a mix collects a certain threshold of packets, decrypts them and shuffles  lexicographically, and finally forwards to the next mix in the cascade. However, the fixed cascade topology scales poorly, since once the cascade reaches its maximum capacity you cannot support more traffic, and this often introduces reliability issues and congestion in packets transmission.
Moreover, the end-to-end latency of packets is unbounded since one cannot predict in which batch their packet will be accepted.
Therefore, although they offered strong anonymity the early deployments of the original Chaumian mixnets have become unfashionable due to a perceived higher latency that cannot accommodate real-time communications.

Over the last two decades, there were significant efforts~\cite{DBLP:conf/sp/DanezisDM03, mixmaster, kwon2019xrd, tyagi2017stadium, van2015vuvuzela, gelernter2016anonpop}
to tackle the scalability, privacy, and latency problems of the early designs anonymous communication systems. In particular, companies like Elixxir, Nym and HOPR focus on developing novel mix network infrastructures to provide anonymous communication. Although all of them develop source-routed decryption mix networks, their key design choices differ regarding topology, mixing technique and onion encryption.

\vspace{2mm}
\textbf{Elixxir}~\cite{elixxir_wp}, a communication infrastracture used by the xx network~\cite{xxnetwork_wp}, implements a variant of the cMix protocol~\cite{DBLP:conf/acns/ChaumDJKKRS17}.  Like the traditional Chaumin mixnet~\cite{DBLP:journals/cacm/Chaum81}, Elixxir uses the \emph{batch-and-reorder} mixing technique.
In order to tackle the problem of time-consuming public key operations and in order to accelerate the packets' processing,
Elixxir introduces a \emph{precomputation} phase,  which allows to perform the public key operations before the \emph{real-time} phase of handling messages between senders and recipients. In result, during the real-time phase, mixes cryptographically process the received packets using the shared values established during the precomputation. However, the time needed to perform the precomputation phase grows linearly with the size of the anonymity set, and it has to be repeated before each real-time communication phase.

The nodes in the network are grouped into small ephemeral \emph{teams}, each one forming a \emph{cascade} topology.
At any given time there might be multiple teams, in varying stages of precomputation phase, only one team is responsible for transporting the network packets in real-time.
Once the batch processing is completed, the team disbands and the nodes can be placed in a new random team. In other words, currently cMix uses a \emph{single cascade} topology. In the future it might be replaced with a \emph{multiple cascade} topology, in which several cascades work in parallel, to handle the increasing traffic.

\vspace{2mm}
\textbf{Nym}~\cite{nym_wp} network is underpinned by the Loopix design~\cite{DBLP:conf/uss/PiotrowskaHEMD17}.
The Nym mixes are grouped into layers which form a \emph{stratified topology}~\cite{DBLP:conf/pet/DingledineSS04}. Each mix in layer $i$ is connected with each mix in the previous and next layer, and traffic flows from the first to the last layer. According to Nym's open-source code base, the current topology has three layers.  In contrast to the original Chaumian mixnet or Elixxir, Nym uses a variant of \emph{continues-time} mixing technique~\cite{kesdogan1998stop}, in which a mix delays each packet before forwarding it to the next hop. The amount of time a packet needs to wait in each mix is chosen by the sender, who picks it at random from an exponential distribution.

In order to circumvent the time-consuming public key operations used for onion encryption, Nym uses the Sphinx cryptographic packet format~\cite{DBLP:conf/sp/DanezisG09}. Sphinx is a compact and efficient cryptographic packet format, that provides bitwise unlinkability for multi-hop routing and supports features like anonymous replies, resistance to tagging attacks, and hiding of the routing information.

\vspace{2mm}
\textbf{HOPR} is building a peer-to-peer based network for anonymous messaging~\cite{hopr_book, hopr_wp}, where participants act both as relay nodes and end users.
Similarly to Nym, packets are encapsulated using Sphinx and routed  through a sequence of few intermediaries to a receiving peer.\footnote{According to the current \href{https://github.com/hoprnet/hoprnet/blob/216de6d1b5f3dcbf99befb06660521c014f567d0/packages/core/src/index.ts\#L719}{code base}, the number of hops is $1$ but can also be user-defined.}
It is not yet clear which mixing technique HOPR peers will use. According to the whitepaper~\cite{hopr_wp}, HOPR is developing a Chaumian mixnet. However, the mixing technique currently implemented works differently from the one used in a Chaumian mix.\footnote{\url{https://github.com/hoprnet/hoprnet/blob/d27a8c506491dc7788259616b9c81307a0ecff97/packages/core/src/mixer.ts\#L25}} Instead of batching and shuffling a fixed number of packets, a HOPR node adds the incoming packets to a single queue from which it picks uniformly at random one packet at the time to process and sends it to the next hop.

\section{Mix network Simulator}\label{sec:simulator}

We developed an open-source mix network simulator using \textsc{Python3}. The interactions between the different network components are simulated using the process-based discrete-event framework \textsc{Simpy}. In this paper, we evaluated our experiments using the
AWS EC2 instances. The code of our simulator is available under an open-source license~\cite{simulator-gitlab}.

\begin{figure*}[t!]
     \centering
     \begin{subfigure}[b]{0.3\textwidth}
         \centering
         \includegraphics[width=\textwidth]{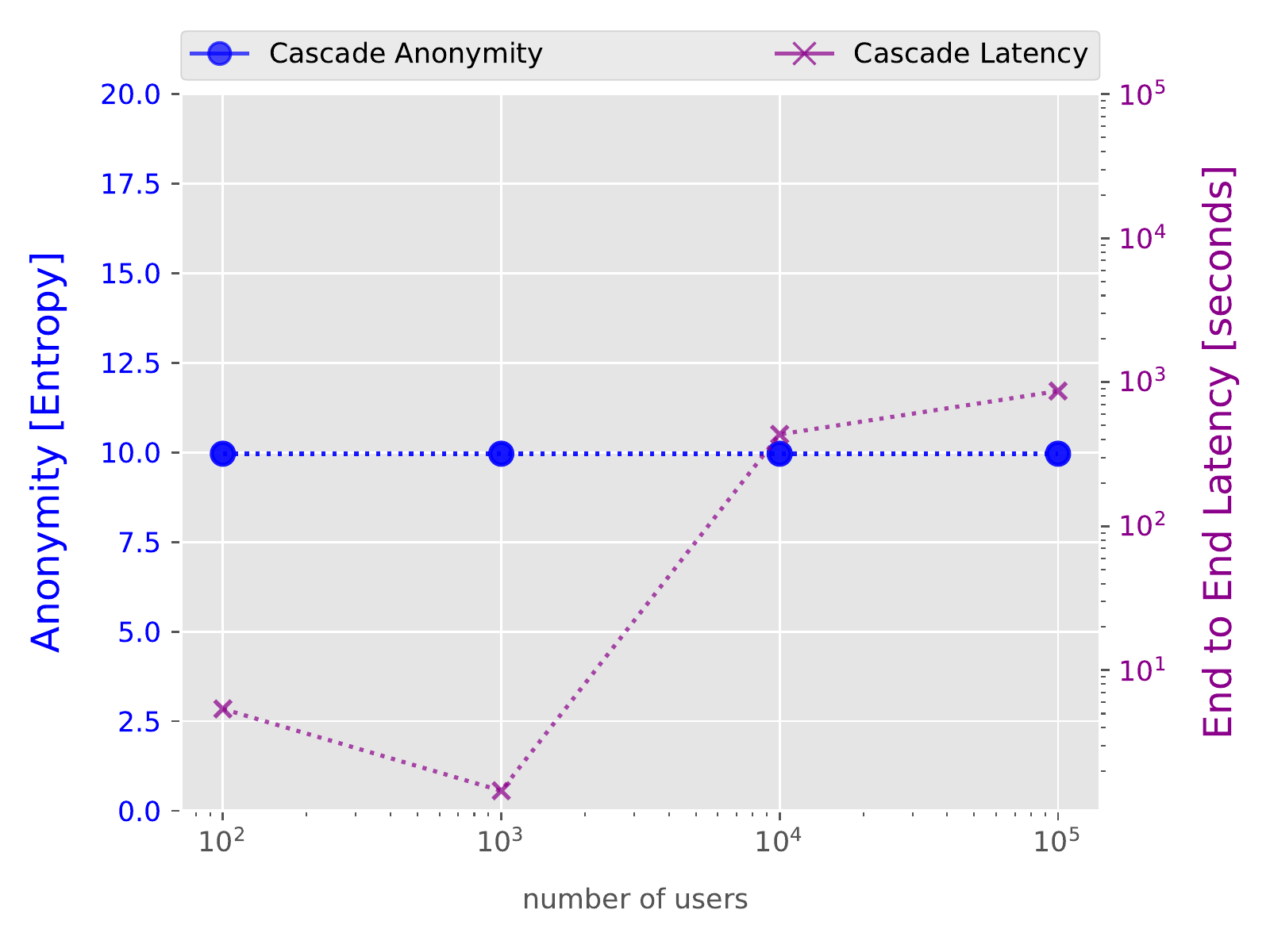}
         \caption{Single cascade Elixxir network}
         \label{fig:cascade}
     \end{subfigure}
     \hfill
     \begin{subfigure}[b]{0.3\textwidth}
         \centering
         \includegraphics[width=\textwidth]{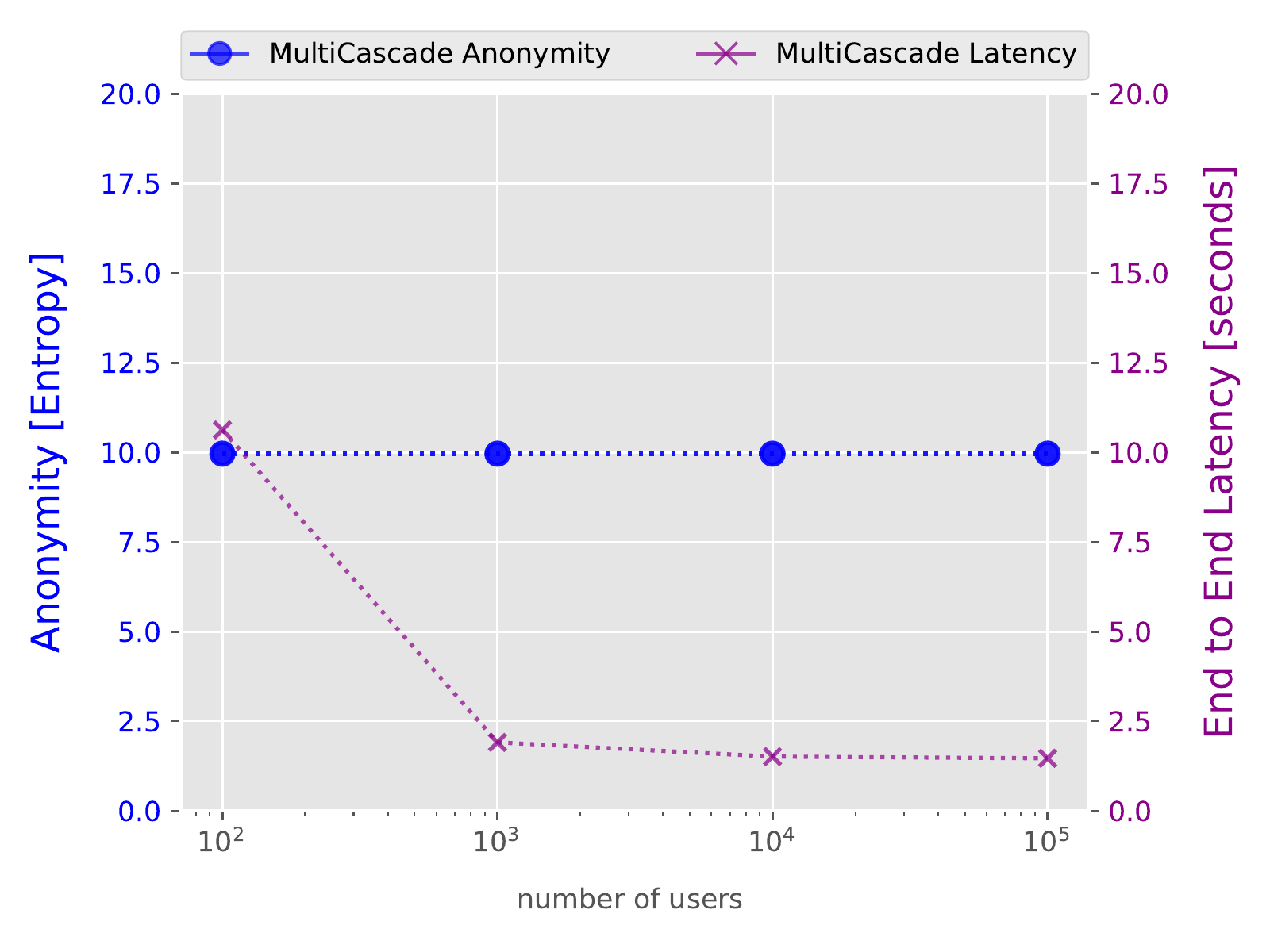}
         \caption{Multi cascade Elixxir network}
         \label{fig:multi_cascade}
     \end{subfigure}
     \hfill
     \begin{subfigure}[b]{0.3\textwidth}
         \centering
         \includegraphics[width=\textwidth]{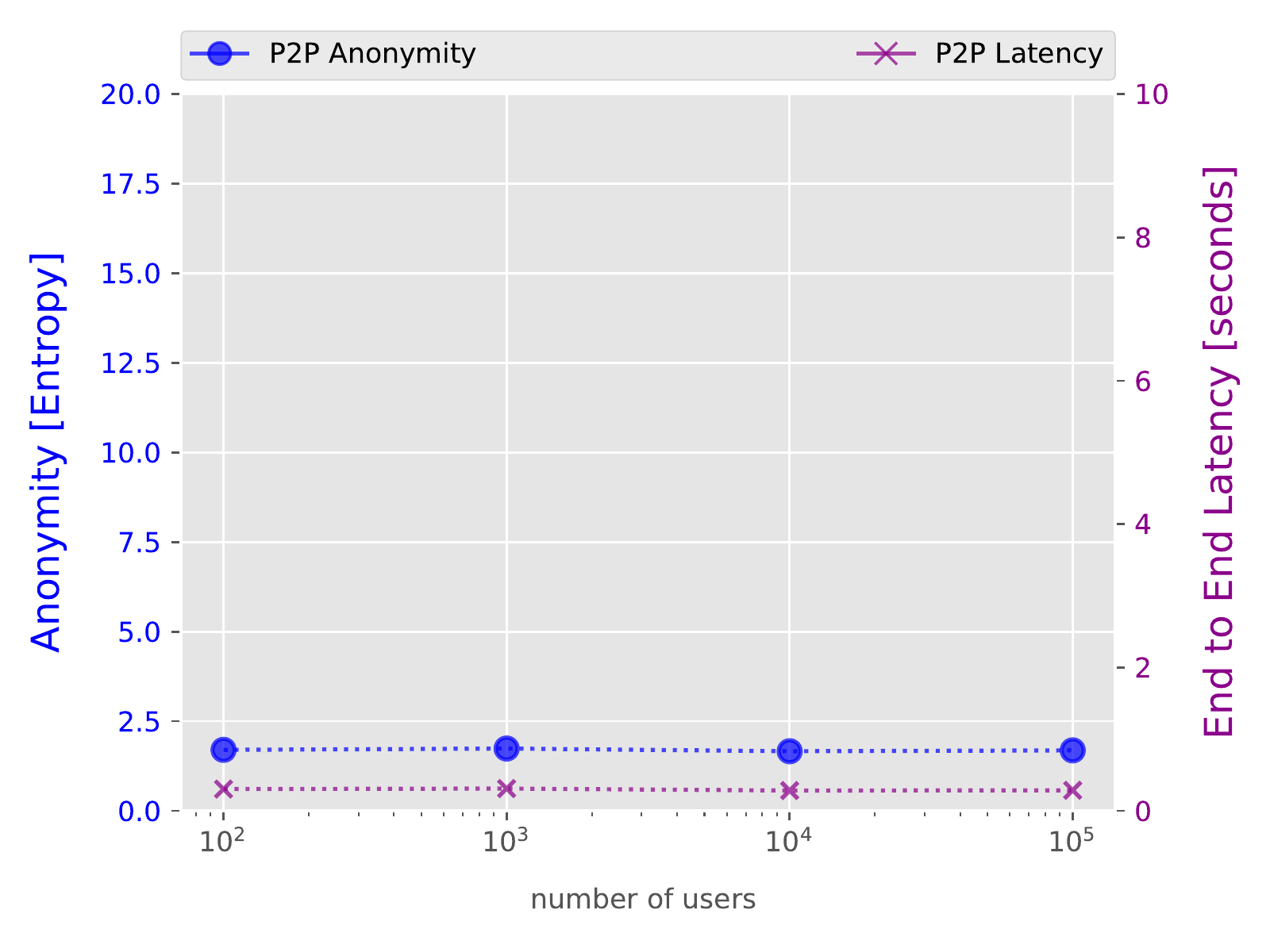}
         \caption{HOPR mix network}
         \label{fig:p2p}
     \end{subfigure}
\end{figure*}

\subsection{Simulator configuration}

\paragraph{Network topology}
Currently, our simulator supports four different network topologies, i.e., (1) cascade, (2) multi-cascade, (3) stratified and (4) peer-to-peer topologies. Further, we also allow for different configuration of the selected topology, including: 
total number of nodes (for cascade and p2p), number of parallel cascades (for multi cascades), and number of layers and nodes per layers (for stratified). 

\paragraph{Mixing technique}
Our simulator implements two mixing techniques, (1) the  \emph{batch and reorder} technique and (2) the \emph{Poisson} mixing technique~\cite{DBLP:conf/uss/PiotrowskaHEMD17}, which is a variant of the continuous time mixing technique~\cite{kesdogan1998stop}. 

\paragraph{Cover traffic} 
An important feature of anonymous communication networks is \emph{cover traffic}, which can be deployed to ensure that there is sufficient traffic in the network to guarantee a large anonymity set, as well as, disguising communication patterns of the individual users. Our implementation, if desired, allows for simulating cover traffic generation by either clients or intermediate nodes. Additionally, the frequency of cover traffic generation can be adjusted accordingly. 

\paragraph{Sending behaviour}
In our simulation, the traffic generated by the end users is modeled by the Poisson distribution. We chose this  distribution as it is appropriate if the arrivals are from a large number of independent sources, like in networks with many clients and nodes. Moreover, Poisson models have been widely used in computer networks and telecommunications literature~\cite{frost1994traffic, heyman2004superposition}. 
The configurable parameter of the Poisson distribution allows simulating different sending patterns of the clients depending on the use case. In addition to the above configurable parameters, our simulator implements also customisable packet and message sizes, and message fragmentation.

\subsection{Anonymity metrics}

We currently implement the two following metrics to quantify privacy offered by the analysed network:

\paragraph{Entropy}

A common measure of anonymity is the \emph{anonymity set}, which reflects the number of other packets with which our message can be confused by the attacker. However, a global passive adversary monitoring the network can assign 
different probabilities for each outgoing packet being linked to the observed incoming packet. This can leak a lot of information to the attacker. 
Therefore, to measure the end to end anonymity we implement the metric based on \emph{Shannon entropy} introduced in~\cite{serjantov2002towards}. In a nutshell, each 
outgoing packet will have a distribution over being linked with past input packets, and the entropy of this distribution is our anonymity metric (see Appendix~\ref{sec:appendix}). By using entropy, we gain a measure of how indistinguishable the packets shuffled by the mix network are among each other, hence how uncertain is the adversary about which of the outgoing packets is his target one.

\paragraph{Third-party unlinkability}
Our simulator implements also the \emph{third-party unlinkability} metric introduced in~\cite{DBLP:conf/uss/PiotrowskaHEMD17}. This metric measures the chances of the adversary correctly correlating the communicating parties even if they know in advance that either $S_1$ or $S_2$ are communicating with $R$. The metric measures the expected difference in likelihood that a given packet received by $R$ is sent from $S_1$ in comparison to $S_2$. 
In Appendix~\ref{sec:appendix} we present the formal definitio of this metric and describe how do we quantify information leaked given multiple rounds of observation.

\section{Empirical study of Elixxir, HOPR and Nym}

\begin{figure}[t!]
    \centering
    \includegraphics[width=\columnwidth]{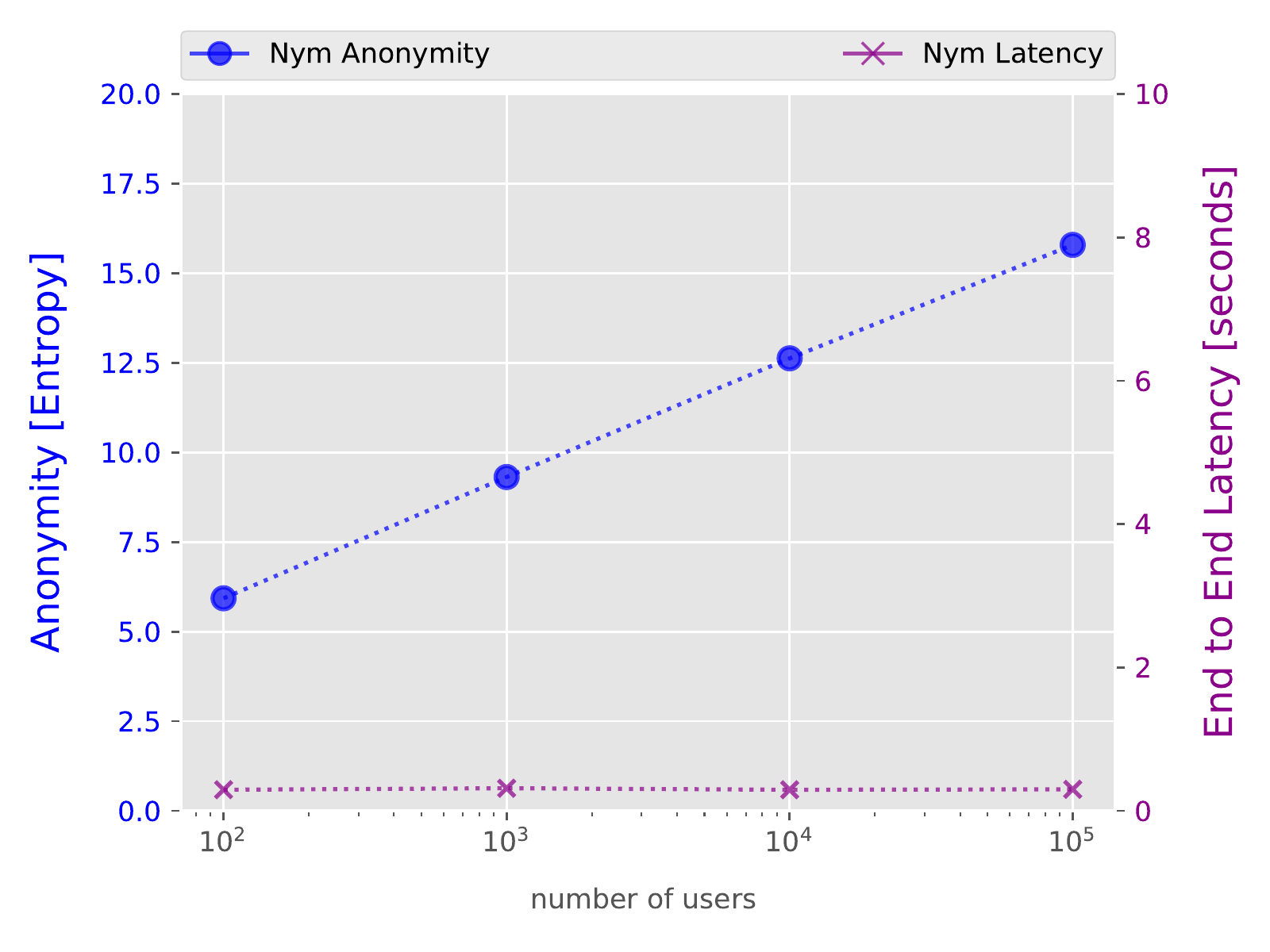}
    \caption{Nym: Impact of increasing user base on anonymity and end to end latency.}
    \label{fig:nym}
\end{figure}

Using the simulator, we analyse the privacy (as measured in by entropy) and performance offered by mix network designs like Elixxir, HOPR and Nym. In particular, we take a closer look at how the core design decisions impact the scalability, anonymity and latency overhead. 

\begin{figure*}[t!]
    \centering
    \begin{minipage}{.48\textwidth}
        \centering
        \includegraphics[width=\textwidth]{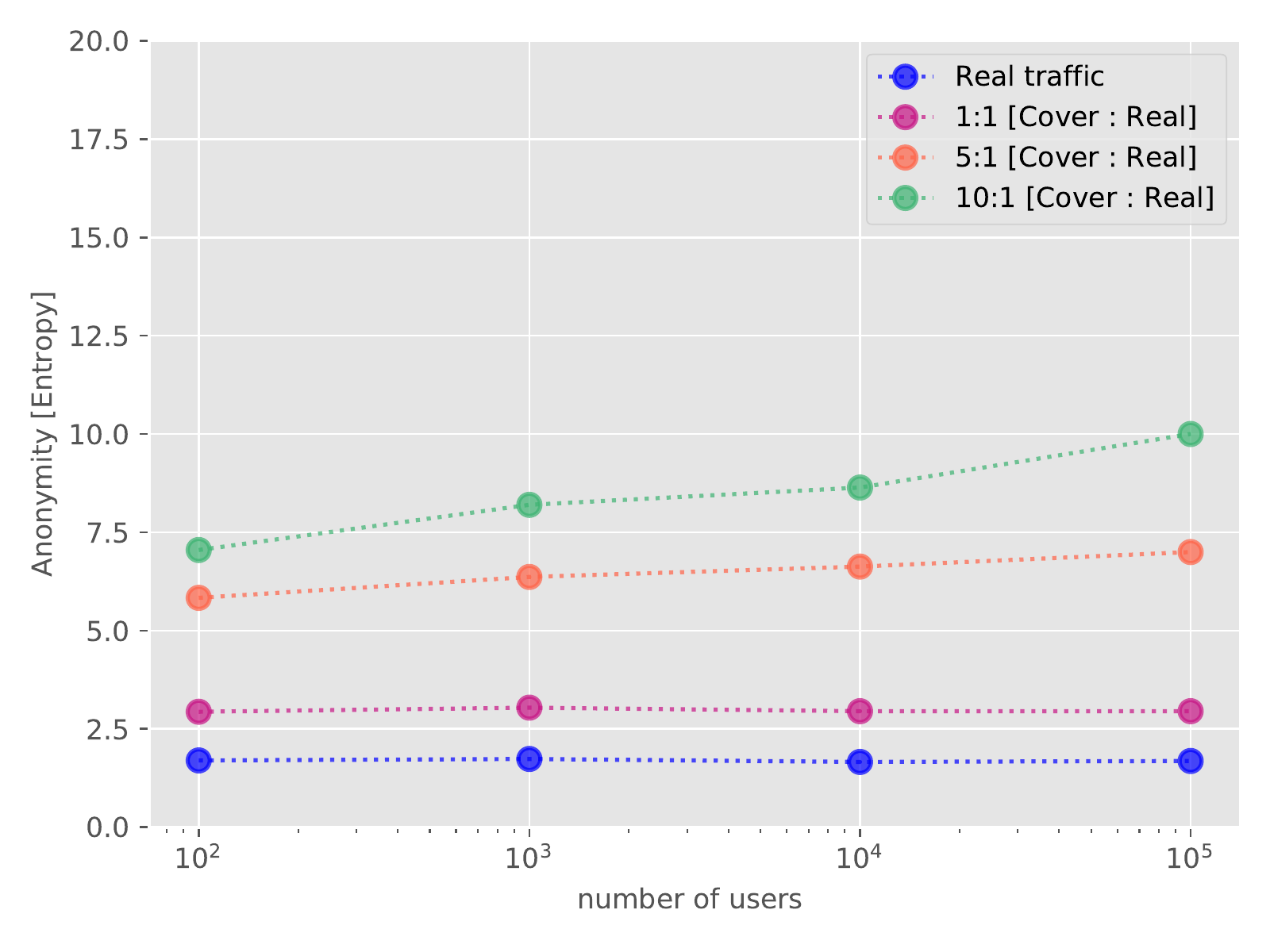}
        \caption{HOPR: Volume of cover traffic needed to provide 
        desired levels of anonymity given increasing user base.}
        \label{fig:p2p_cover}
    \end{minipage}\hfill
    \begin{minipage}{.48\textwidth}
        \centering
        \includegraphics[width=\textwidth]{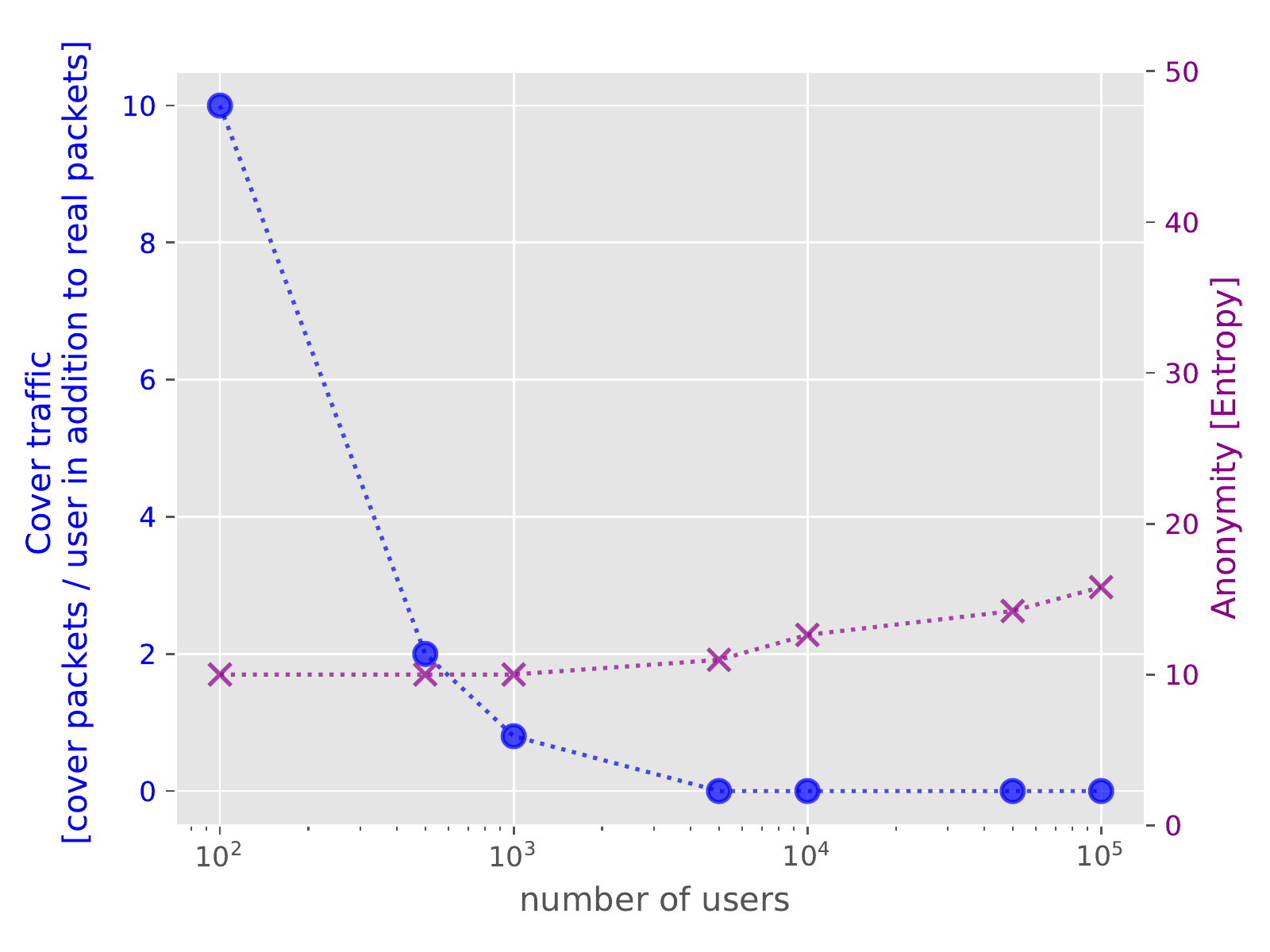}
        \caption{Nym: Volume of cover traffic needed to provide 
        minimum desired anonymity $10$ in entropy given increasing user base.}
        \label{fig:users_vs_cover}
    \end{minipage}\hfill
\end{figure*}

\paragraph{Assumptions} 
In order to make our analysis fair, we make the following assumptions. 
For the HOPR and Nym networks we assume the same processing capacity of the mix nodes (i.e., single node has a capacity to process $1000$ packets per second). Furthermore, in the case of Elixxir network we assume a batch of size $1000$ packets.
Moreover, we exclude from our analysis in terms of end-to-end latency the time needed for the cryptographic processing of the packets, as this time depends on the underlying onion encryption protocol, as well as the choice of the implementation language and the optimization of the code base. The end-to-end latency is given purely by the time needed to mix the packets.

Although HOPR and Nym implement different mixing techniques (see Section~\ref{sec:background}), from a theoretical standpoint the information leaked to the adversary observing a single mix node is the same. This is due to the fact that the exponential distribution used by Nym to generate the intermediate delays has the \emph{memoryless property}~\cite{mitzenmacher2005probability}, meaning that the probability that a packet leaves a mix at a given time is independent of its arrival time. Thus, from the perspective of the network observer each packet in the mix's pool has equal chances to be the next one to be forwarded. Therefore, in our simulation we use for both HOPR and Nym the Poisson mixing technique~\cite{DBLP:conf/uss/PiotrowskaHEMD17}. Without a loss of generality, unless stated otherwise, we assume the mean parameter of the exponential distribution to be $0.1$ second per node. Similarly, we assume that each end user sends on average $1$ packet per second following a Poisson process. Finally, all packets between end users are routed via $3$ mix nodes. 

\vspace{1mm}
\emph{Results.} Anonymity loves company~\cite{DBLP:conf/weis/DingledineM06}, hence scalability is one of the key properties of any anonymous communication network. 
As shown in Figure~\ref{fig:cascade} mix networks which have a cascade topology and batch-and-reorder mixing techniques (like Elixxir) scale poorly with an increasing user base. This design imposes large latency overhead when the volume of traffic grows, as the cascade becomes congested and each packet has to wait for a long period of time to be processed. Note that this design imposes high latency overhead not only when the volume of traffic is high but also when the traffic volume is low since each node waits a long time to collect a full batch for processing. Thus, end users have no control or prediction on the end-to-end latency of their packets. 

An alternative approach, which significantly improves scalability is to introduce multiple cascades which work in parallel, a solution proposed by Elixxir.\footnote{A new cascade is added only once all of the existing ones are at their full capacity.} 
As shown in Figure~\ref{fig:multi_cascade}, such approach allows indeed the network to scale out and thus avoid a high latency overhead. However, the anonymity provided is still always constant as it is limited to the fixed size of the batch. Thus the mix network does not profit from the increasing number of users. Moreover, even with multiple cascades,  Elixxir still imposes much higher end-to-end latency that HOPR and Nym. 
As Figure~\ref{fig:p2p} shows, networks like HOPR that 
 combine a P2P topology with continuous-time mixing scale well but provide very poor anonymity even when there is an increasing number of users, because the user traffic is spread out thinly through the network.  
Moreover, since HOPR nodes select  packets to forward randomly from the pool of currently stored packets, the end user has no guarantees on bounds for delivery time of sent packets, as is also the case of the batch and reorder technique.

\begin{figure}[t!]
    \centering
    \includegraphics[width=\columnwidth]{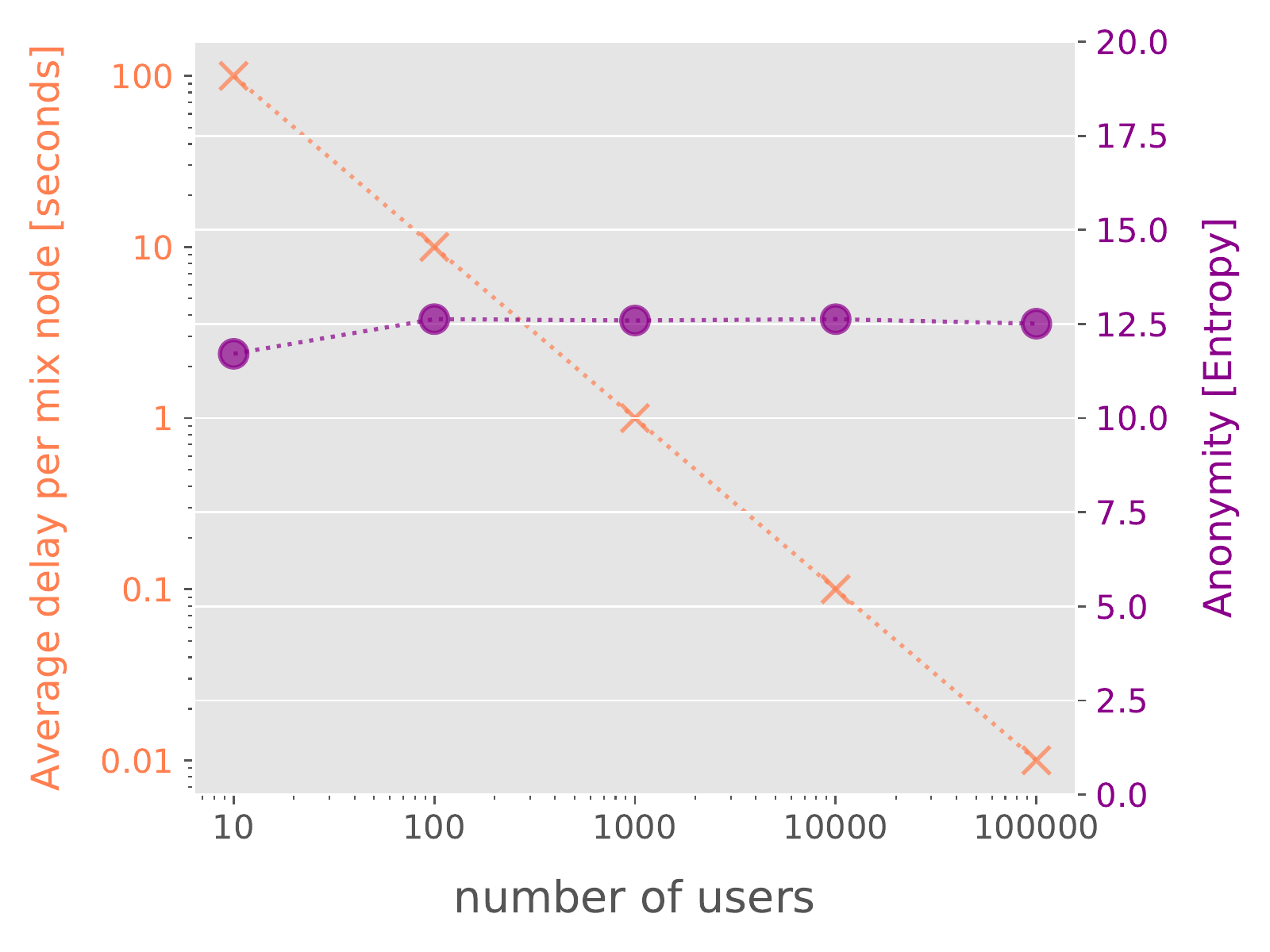}
    \caption{Nym: Impact of increasing volume of traffic on required mixing delay.}
    \label{fig:users_vs_delay}
\end{figure}

On the other hand, as shown on Figure~\ref{fig:nym}, the stratified topology used by Nym scales out by adding more nodes per layer to support a large user base without a negative impact on the end-to-end latency. In particular, the end to end latency is bound by the mean parameter of the exponential distribution used by the individual senders.\footnote{In implementation, it would also be bound by the time of processing the cryptographic packet format used for encryption.} 
Critically, the anonymity provided by Nym increases as the network grows. The more traffic flows into the Nym mix network, the larger the anonymity as measured in entropy. This is due to the fact, that although the traffic flows via multiple independent routes, the connectivity of a stratified topology guarantees that the routes intersect, hence all packets contribute to the same anonymity set. In contrast to the multi-cascade Elixxir and HOPR, Nym provides much better anonymity for large volumes of users. What differentiates Nym from designs like Elixxir and HOPR is that the increasing number of Nym users allows to tune down the mean parameter of the exponential distribution, and thus decrease the end-to-end latency, without sacrificing anonymity (Figure~\ref{fig:users_vs_delay}).

One way to improve the level of anonymity is to deploy cover traffic. However, as our analysis shows (Figure~\ref{fig:p2p_cover}), P2P networks like HOPR need huge volumes of cover traffic to reach reasonable levels of anonymity. In contrast, as Figure~\ref{fig:users_vs_cover} depicts, once the number of users increases in the Nym mix network (and hence the volume of real packets flowing into the mix network), the amount of cover traffic needed to provide a constant desired anonymity decreases. Moreover, in contrast to HOPR in which the end users carry the burden (and cost) of generating the cover traffic, in Nym this responsibility can be delegated to the mixes. 




\section{Conclusion}\label{sec:conclusion}

In this paper, we outlined the design and implementation of a novel discrete-event mix network simulator, which can be used to investigate how different network parameters, design choices or user loads impact both the anonymity as well as the latency and the bandwidth overhead. We further use our simulator as a tool to compare three fast growing mix network infrastructure: Elixxir, HOPR and Nym.

\bibliographystyle{ACM-Reference-Format}
\bibliography{mybib}

\appendix

\section{A direct name for the appendix}\label{sec:appendix}
\subsection{Shannon Entropy}
Let $X$ be a discrete random variable over the finite set $\mathcal{X}$ with probability mass function $p(x) = Pr(X = x)$. The Shannon entropy $H(X)$ of a discrete random variable $X$ is defined as 
\begin{align*}
    H(X) = - \sum_{x \in \mathcal{X}} p(x) \log{p(x)}.
\end{align*}

\subsection{Sender-Receiver Unlinkability}

The \emph{sender-receiver unlinkability} metric~\cite{DBLP:conf/uss/PiotrowskaHEMD17} is defined as follows. 
Let $I_1$ and $I_2$ denote events that the sender of the target packet was $S_1$ or $S_2$ respectively. 
As $\epsilon$ we denote the measure of information leaked by the system and define it as
\begin{align*}
    \epsilon = \log \left( \frac{\Pr(I_1)}{\Pr(I_2)} \right)
\end{align*}

$\epsilon$ is the maximum leakage the adversary can learn from observing the system during events $I_1$ and $I_2$. 
Additionally, we denote as $\delta$ the probability by which the leakage exceeds this $\epsilon$.

\vspace{2mm}
The above definition quantifies the information leakage in a single round of observation. However, multiple observations  increase the adversary’s advantage. 
One way to compute the leakage for multiple rounds is to apply the \emph{differential privacy composition theorem}~\cite{Dwork}. However, this theorem assumes the worst case leakage in each round. Instead, we 
compute the average case leakage. 
From the Law of Large numbers~\cite{feller1968introduction}, we know that for large number of repeated experiments the average of the results converges to the expected value of the outcomes. Hence, in order to compute the anonymity leakage of an observed round $R_i$ we compute the value of $\hat{\epsilon}$ for a large number of samples $K$ as follows
\begin{align*}
  \hat{\epsilon} = \frac{1}{K} \sum_{i=1}^K \log \left(\frac{\Pr_i(I_1)}{\Pr_i(I_2)}\right).
\end{align*}

The overall advantage after $R$ observations can be expressed as $\epsilon_R = R\cdot \hat{\epsilon}$.

\end{document}